\documentclass[aps,prd,twocolumn,groupedaddress,preprintnumbers,nofootinbib]{revtex4-2}

\usepackage{amssymb}
\usepackage{amsmath}
\usepackage{graphicx}
\usepackage{enumerate}
\usepackage{mathtools}
\usepackage{color}
\usepackage{bbold}
\usepackage{subfigure}
\usepackage{slashed}
\usepackage[dvipsnames,usenames]{xcolor}
\usepackage{bm}
\usepackage{bbm}
\usepackage{soul}
\usepackage{centernot}

\begin{document}


\preprint{LA-UR-24-20772}

\title{Subgrid modeling of neutrino oscillations in astrophysics}



\author{Lucas Johns}
\email[]{ljohns@lanl.gov}
\affiliation{Theoretical Division, Los Alamos National Laboratory, Los Alamos, NM 87545, USA}

\begin{abstract}
Approximating neutrino oscillations as subgrid physics is an appealing prospect for simulators of core-collapse supernovae and neutron star mergers. Because flavor instabilities quickly lead to quasisteady states in oscillation calculations, it is widely believed that flavor mixing can be approximated in astrophysical simulations by mapping unstable states onto the appropriate asymptotic ones. Subgrid models of this kind, however, are not self-consistent. The miscidynamic theory of quantum-coherent gases furnishes a subgrid model that is.
\end{abstract}

\maketitle

\section{Introduction\label{sec:intro}}

Integrating neutrino oscillations into the theory of core-collapse supernovae and neutron star mergers is a longstanding challenge at the frontiers of particle physics, astrophysics, and quantum statistical physics. Efforts to solve the oscillation problem cluster into three paradigms:

\textbf{(1) Solving the quantum kinetic equation.} Neutrino quantum kinetics is a well-established theory whose governing equation is the subject of intense study:
\begin{equation}
i \left( \partial_t + \boldsymbol{\hat{p}} \cdot \partial_{\boldsymbol{x}} \right) \rho_{\boldsymbol{p}} (t, \boldsymbol{x}) = \left[ H_{\boldsymbol{p}} (t, \boldsymbol{x}), \rho_{\boldsymbol{p}} (t, \boldsymbol{x}) \right] + i C_{\boldsymbol{p}} (t, \boldsymbol{x}), \label{eq:qke}
\end{equation}
with Hamiltonian $H_{\boldsymbol{p}}$ and collision term $iC_{\boldsymbol{p}}$, both of which are functionals of the background particle densities and the neutrino density matrices $\rho_{\boldsymbol{q}}$ at all momenta $\boldsymbol{q}$ \cite{fuller1987, notzold1988, pantaleone1992, sigl1993, raffelt1993, raffelt1993b, loreti1994, qian1995a, sigl1995, pantaleone1995, yamada2000, pastor2002b, friedland2003, strack2005, balantekin2005, cardall2008, volpe2013, vlasenko2014, kartavtsev2015, stirner2018, richers2019}. Due to computational cost, it is not possible to solve this equation in a realistic compact-object model. The astrophysical environment must instead be treated in a drastically simplified manner \cite{duan2010, dasgupta2012, stapleford2020, shalgar2023e, nagakura2023b, nagakura2023e}. Toy-model quantum-kinetic calculations continue to deepen our understanding of oscillation phenomenology. However, as a way of solving the oscillation problem, this program---direct numerical simulation of neutrino quantum kinetics---is severely hampered by the apparent lack of models that are both physically adequate and computationally practical. Specific tactics like angular-moment truncations \cite{duan2014, johns2020, johns2020b, myers2022, grohs2023, grohs2023b} and artificial attenuation \cite{nagakura2022b} are actively being explored, but currently it is unclear whether they can salvage the broader approach.

\textbf{(2) Using a subgrid model motivated by small-scale oscillation calculations.} The possibility has long been contemplated that flavor instabilities result in asymptotic quasisteady states in compact-object environments \cite{sawyer2005}. With fast instabilities being widespread in state-of-the-art simulations \cite{wu2017, abbar2021b, nagakura2021c, akaho2023}, the belief that this does indeed occur has become the basis of a distinct paradigm around which research has been coalescing: the goal of oscillation research is not to simulate quantum kinetics but rather to supply accurate mappings from pre- to post-instability neutrino distributions. Rapid progress is being made in this direction \cite{bhattacharyya2020, richers2021b, bhattacharyya2021, richers2021c, wu2021, richers2022d, bhattacharyya2022, zaizen2023, zaizen2023b, xiong2023d, abbar2023}. Just as importantly, studies are demonstrating the feasibility of implementing asymptotic-state subgrid models within astrophysical simulations \cite{li2021, padillagay2021b, just2022, fernandez2022, ehring2023b, ehring2023c}.

\textbf{(3) Solving the miscidynamic equation.} An alternative proposal is to simulate hydrodynamics coupled to neutrino miscidynamics, a coarse-grained transport theory that follows from applying a local-equilibrium approximation to neutrino oscillations \cite{johns2023c}. The prefix \textit{misc-}\footnote{Pronounced with a soft \textit{c} as in \textit{miscible} or \textit{inviscid}.} refers to local \textit{mixing} equilibrium. The (adiabatic) miscidynamic equation is
\begin{equation}
i \left( \partial_t + \boldsymbol{\hat{p}} \cdot \partial_{\boldsymbol{x}} \right) \rho^\textrm{eq}_{\boldsymbol{p}} (t, \boldsymbol{x}) = i C^\textrm{eq}_{\boldsymbol{p},\textrm{non}} (t, \boldsymbol{x}), \label{eq:misc}
\end{equation}
where $\rho_{\boldsymbol{p}}^\textrm{eq}$ is the mixing-equilibrium distribution. The collisional term $i C_{\boldsymbol{p}}^\textrm{eq}$ is the same as in Eq.~\eqref{eq:qke} but with the replacement $\rho_{\boldsymbol{p}} \rightarrow \rho_{\boldsymbol{p}}^\textrm{eq}$ and with only the instantaneously non-unitary part retained. Although approaches \textbf{(2)} and \textbf{(3)} both advocate subgrid modeling, miscidynamics entails a radically different vision of neutrino oscillations in supernovae and mergers. It proposes, for example, that flavor instabilities in these sites may not take the form previously expected.

In this paper we identify a logical inconsistency in \textbf{(2)}. The point is simple, but it calls for a redirection in the research area aimed at solving the oscillation problem. In fact, consideration of how to resolve the self-consistency issue leads in the direction of \textbf{(3)}. The miscidynamic approach to flavor transport parallels the use of hydrodynamics for the description of momentum transport. It is rooted in compelling physical principles, is systematically improvable, and has a hope of being computationally feasible.

Before presenting these points, let us take the opportunity to make one other simple but consequential remark. Asymptotic-state subgrid models are in many cases solely concerned with fast flavor instabilities. Collisional flavor instabilities \cite{johns2023, johns2022b, lin2023, padillagay2022b, xiong2023, liu2023, xiong2023c, nagakura2023c, shalgar2023b, liu2023b, fischer2023, kato2023b, fiorillo2023d, johns2023d}, the Mikheyev--Smirnov--Wolfenstein (MSW) effect \cite{wolfenstein1978, mikheyev1985}, spectral swaps/splits \cite{duan2006, duan2006b, duan2006c, duan2007, duan2007b, duan2007c, raffelt2007, raffelt2007c, fogli2007, fogli2008, fogli2009b, dasgupta2008c, dasgupta2009, dasgupta2010, friedland2010, galais2011}, and matter--neutrino resonances \cite{malkus2012, malkus2014, malkus2016, wu2016, vaananen2016, zhu2016} involve the grid-scale development of flavor coherence and are incompatible with most asymptotic-state proposals. Miscidynamics has the potential to accommodate all oscillation phenomena, including those just listed.

\section{Self-consistency\label{sec:issue}}

Fast flavor instabilities lead to asymptotic quasi-steady states in oscillation calculations. As a result of this process, unstable conditions are erased on oscillation time scales ($\lesssim 1~\mu$s). Unstable conditions are tied to astrophysical features like the fluid density profile and therefore emerge on astrophysical time scales ($\gtrsim 1$~ms).

Motivated by the large scale disparity, subgrid models have assumed that the time to reach asymptotic states can be ignored, \textit{i.e.}, that asymptotic states are reached infinitely quickly. In this limit, instabilities lead to the erasure of unstable conditions instantaneously. But this implies that instabilities can never develop in the first place. Thus asymptotic-state subgrid models are self-contradictory.

To make the argument more concrete, consider a relaxation-time approximation for the evolution toward the local asymptotic state $\rho^\textrm{a}_{\boldsymbol{p}} (t, \boldsymbol{x})$ \cite{nagakura2023d}: 
\begin{align}
i \left( \partial_t + \boldsymbol{\hat{p}} \cdot \partial_{\boldsymbol{x}} \right) \rho_{\boldsymbol{p}} (t, \boldsymbol{x}) = \gamma (t, \boldsymbol{x}) &\left( \rho^\textrm{a}_{\boldsymbol{p}} (t, \boldsymbol{x}) - \rho_{\boldsymbol{p}} (t, \boldsymbol{x}) \right) \notag \\
&~~+ i C_{\boldsymbol{p}} (t, \boldsymbol{x}). \label{eq:bgk}
\end{align}
Asymptotic-state subgrid models assume that the relaxation time $\tau$ (inverse relaxation rate $\gamma^{-1}$) is much smaller than the linear grid dimension,
\begin{equation}
\tau (t, \boldsymbol{x}) \equiv \gamma^{-1} (t, \boldsymbol{x}) \ll l_\textrm{astro} (t, \boldsymbol{x}). \label{eq:gamma}
\end{equation}
In fact, because the transient evolution leading to the asymptotic state is neglected entirely, these models correspond to the assumption that flavor relaxation is instantaneous:
\begin{equation}
\tau (t, \boldsymbol{x}) \cong 0. \label{eq:tau}
\end{equation}
From this it follows that
\begin{equation}
\rho_{\boldsymbol{p}} (t, \boldsymbol{x}) \cong \rho^\textrm{a}_{\boldsymbol{p}} (t, \boldsymbol{x}). \label{eq:rhoa}
\end{equation}

The general routine in asymptotic-state subgrid models consists of two steps. The first advances the simulation forward without oscillations, \textit{i.e}, solves the Boltzmann equation
\begin{equation}
\left( \partial_t + \boldsymbol{\hat{p}} \cdot \partial_{\boldsymbol{x}} \right) f_{\alpha, \boldsymbol{p}} (t, \boldsymbol{x}) = C_{\alpha, \boldsymbol{p}} (t, \boldsymbol{x}) \label{eq:boltz}
\end{equation}
for the neutrino distribution function $f_{\alpha, \boldsymbol{p}}$ of each flavor $\alpha$. We use $C_{\alpha, \boldsymbol{p}}$ to denote the collision term without flavor coherence and ignore general-relativistic effects for simplicity of presentation. The second step then imposes Eq.~\eqref{eq:rhoa}, where $\rho^\textrm{a}_{\boldsymbol{p}} (t, \boldsymbol{x})$ is the asymptotic state resulting from the initial state obtained at the end of the first step:
\begin{equation}
\left[ \rho^\textrm{i}_{\boldsymbol{p}} (t, \boldsymbol{x}) \right]_{\alpha\beta} = \begin{cases}
f_{\alpha, \boldsymbol{p}} (t, \boldsymbol{x}) & \alpha = \beta \\
0 & \alpha \neq \beta
\end{cases}
\end{equation}
The \textit{initial} label refers to the fact that $\rho^\textrm{a}_{\boldsymbol{p}}$ is calculated using $\rho^\textrm{i}_{\boldsymbol{p}}$ as the initial conditions (up to small perturbations) in an oscillation calculation. Paradigm $\textbf{(2)}$ makes this $\rho^\textrm{i}_{\boldsymbol{p}} \rightarrow \rho^\textrm{a}_{\boldsymbol{p}}$ mapping the central concern of the oscillation problem.

Is this alternation of steps a reasonable approximation of evolution under the condition in Eq.~\eqref{eq:rhoa}? No, not in general. As neutrinos move from cell $\boldsymbol{x}_1$ at $t_1$ to cell $\boldsymbol{x}_2$ at $t_2$, they violate Eq.~\eqref{eq:rhoa} because this step is carried out by solving Eq.~\eqref{eq:boltz}, which effectively forbids neutrinos from oscillating. Turning off oscillations is only justified for distances smaller than the oscillation length $l_\textrm{osc}$, which in turn is smaller than $\tau$. In the routine described above, the distance over which they are turned off is $l_{\textrm{astro}}$. But then this method is only justified if the simulation resolves cells of size
\begin{equation}
l_\textrm{astro} (t, \boldsymbol{x}) \lesssim \tau (t, \boldsymbol{x}) \cong 0.
\end{equation}
The condition [Eq.~\eqref{eq:gamma}] that is meant to justify asymptotic-state subgrid models is the very condition that makes them questionable.

Note that Eq.~\eqref{eq:rhoa} is equivalent to the statement of local mixing equilibrium upon identifying
\begin{equation}
\rho^\textrm{a}_{\boldsymbol{p}} (t, \boldsymbol{x}) = \rho^\textrm{eq}_{\boldsymbol{p}} (t, \boldsymbol{x}).
\end{equation}
Unlike asymptotic-state subgrid models, miscidynamics self-consistently imposes Eq.~\eqref{eq:gamma} in approximating neutrino quantum kinetics. We discuss this point further in the next section.

Let us also contrast asymptotic-state subgrid models with a technique used successfully in calculations of neutrino oscillations during the cosmological epoch of weak decoupling (\textit{e.g.}, Ref.~\cite{froustey2020}). In the early-universe context, it is common to perform an operator splitting in which oscillations are advanced with a step size $t_\textrm{osc}$ and collisions with a step size $t_\textrm{coll} \gg t_\textrm{osc}$. During the oscillation (collision) steps, the collision (oscillation) term is dropped from the quantum kinetic equation. However, oscillations are not entirely neglected during the long collision steps because collision integrals are computed as functionals of the density matrices $\langle \rho^\textrm{osc}_{\boldsymbol{p}} \rangle$, where $\rho^\textrm{osc}_{\boldsymbol{p}}$ is the solution of the oscillation-only evolution and angle brackets average over duration $t_\textrm{coll}$. The quantities $\langle \rho^\textrm{osc}_{\boldsymbol{p}} \rangle$ are analogous to $\rho^\textrm{a}_{\boldsymbol{p}}$, but oscillations are effectively kept on throughout the long collisional steps, as demanded by the scales in the problem.

The issue with asymptotic-state subgrid models is that they inconsistently apply scale separation. If a neutrino system responds quickly \textit{after} a change in parameters (\textit{e.g.}, the neutrino momentum distributions, the electron density), it should also respond quickly \textit{during} the change. Self-consistency requires consideration of how neutrinos move between cells in a simulation. Oscillations cannot simply be turned off during intercell transit.

The inability to isolate some part of a supernova or merger from its neighboring regions is a signature challenge of the oscillation problem. Subgrid models, though prescribing outcomes at small scales, must be informed by a global view of the physics. Similar points were made in Ref.~\cite{johns2022} regarding the effects of collisions and Ref.~\cite{johns2023b} regarding the effects of quantum entanglement. Reflecting on how simple calculations are meant to fit into the full problem is a constant necessity in this research area.

\section{Miscidynamics\label{sec:misc}}

Rather than formulating adiabatic miscidynamics on the basis of thermodynamic principles, as done in Ref.~\cite{johns2023c}, we will here derive it---and infer $\rho^\textrm{eq}_{\boldsymbol{p}}$---directly from the principles ostensibly underlying paradigm $\textbf{(2)}$. We will thus show how the mandate of self-consistency in fact leads to $\textbf{(3)}$.

We take Eq.~\eqref{eq:bgk} as our starting point and replace $\rho^\textrm{a}_{\boldsymbol{p}} (t, \boldsymbol{x})$ with $\rho^\textrm{eq}_{\boldsymbol{p}} (t, \boldsymbol{x})$:
\begin{align}
i \left( \partial_t + \boldsymbol{\hat{p}} \cdot \partial_{\boldsymbol{x}} \right) \rho_{\boldsymbol{p}} (t, \boldsymbol{x}) = \gamma (t, \boldsymbol{x}) &\left( \rho^\textrm{eq}_{\boldsymbol{p}} (t, \boldsymbol{x}) - \rho_{\boldsymbol{p}} (t, \boldsymbol{x}) \right) \notag \\
&~~+ i C_{\boldsymbol{p}} (t, \boldsymbol{x}). \label{eq:bgkeq}
\end{align}
At this stage we are not specifying precise meanings of $\rho^\textrm{eq}_{\boldsymbol{p}}$ and $\gamma$. We are merely assuming that the effect of the oscillation term $\left[ H_{\boldsymbol{p}} (t, \boldsymbol{x}), \rho_{\boldsymbol{p}} (t, \boldsymbol{x}) \right]$ is to bring $\rho_{\boldsymbol{p}} (t, \boldsymbol{x})$ to some state $\rho^\textrm{eq}_{\boldsymbol{p}} (t, \boldsymbol{x})$ at some rate $\gamma (t, \boldsymbol{x})$. We further assume, as asymptotic-state subgrid models do, that finite-$\tau$ dynamics is negligible and  $\rho^{\textrm{eq}}_{\boldsymbol{p}}$ is spatially homogeneous on subgrid scales. Under these assumptions, oscillations have no effect other than to make $\rho_{\boldsymbol{p}}$ cling to $\rho^{\textrm{eq}}_{\boldsymbol{p}}$. Let us adopt the local equilibrium frame in which $\rho^{\textrm{eq}}_{\boldsymbol{p}}$ is diagonalized. We then have
\begin{equation}
i \left( \partial_t + \boldsymbol{\hat{p}} \cdot \partial_{\boldsymbol{x}} \right) \rho^\textrm{eq}_{\boldsymbol{p}} (t, \boldsymbol{x}) = i C^\textrm{eq}_{\boldsymbol{p},\textrm{non}} (t, \boldsymbol{x}). \label{eq:notmisc}
\end{equation}
Here $i C^\textrm{eq}_{\boldsymbol{p},\textrm{non}}$ is the collision term $i C^\textrm{eq}_{\boldsymbol{p}}$ with the off-diagonal components zeroed out. The off-diagonal parts are absorbed into the frame transformation. If this were not the case, collisions would cause $\rho_{\boldsymbol{p}}$ to develop a finite deviation from $\rho^{\textrm{eq}}_{\boldsymbol{p}}$, contradicting the assumption of instantaneous relaxation. The grid-level effect of the $\gamma$ relaxation term is similarly absorbed into the frame transformation.

Equation~\eqref{eq:notmisc} is formally identical to the miscidynamic equation [Eq.~\eqref{eq:misc}], but we will refrain from calling it such until we have established that it is fully consistent with the theory of Ref.~\cite{johns2023c}. When we refer to Eq.~\eqref{eq:notmisc} rather than Eq.~\eqref{eq:misc}, we mean the $\tau \rightarrow 0$ limit of Eq.~\eqref{eq:bgkeq} with homogeneous asymptotic states. We use the same symbol $\rho^\textrm{eq}_{\boldsymbol{p}}$ because ultimately we will show that Eqs.~\eqref{eq:misc} and \eqref{eq:notmisc} are substantively as well as formally identical.

To accomplish this, we consider the implications of self-consistency. The principle has a global and a local form. \textit{Global self-consistency} was the subject of Sec.~\ref{sec:issue}. It concerns the relationship between flavor distributions at different points in time and/or space. Suppose we have a solution $\rho_{\boldsymbol{p}} (t, \boldsymbol{x})$ to Eq.~\eqref{eq:notmisc}. Then the solutions $\rho_{\boldsymbol{p}} (t_1, \boldsymbol{x}_1)$ and $\rho_{\boldsymbol{p}} (t_2, \boldsymbol{x}_2)$ at any two points are connected by a sequence of $\rho^\textrm{eq}_{\boldsymbol{p}}$ distributions. Being a solution to Eq.~\eqref{eq:notmisc}, $\rho_{\boldsymbol{p}} (t, \boldsymbol{x})$ varies on an astrophysical scale,
\begin{equation}
\partial_{\boldsymbol{x}} \rho_{\boldsymbol{p}} (t, \boldsymbol{x}) \sim l_\textrm{astro}^{-1} (t, \boldsymbol{x}). \label{eq:lastro}
\end{equation}
Here $l_\textrm{astro}$ is a function of neutrino advection [the left-hand side of Eq.~\eqref{eq:notmisc}] and collisions [the right-hand side of Eq.~\eqref{eq:notmisc}]. It may coincide with the grid resolution of a simulation as before. Since the oscillation term drops out everywhere, solution $\rho_{\boldsymbol{p}} (t, \boldsymbol{x})$ has no dependence on the oscillation scale $l_\textrm{osc}$.

\textit{Local self-consistency} is the requirement that a solution of Eq.~\eqref{eq:notmisc} be compatible with the quantum kinetic equation [Eq.~\eqref{eq:qke}]. In the local equilibrium frame, we require
\begin{equation}
\left[ H_{\boldsymbol{p}} (t, \boldsymbol{x}), \rho_{\boldsymbol{p}} (t, \boldsymbol{x}) \right] = \left[ H^\textrm{eq}_{\boldsymbol{p}} (t, \boldsymbol{x}), \rho^\textrm{eq}_{\boldsymbol{p}} (t, \boldsymbol{x}) \right] = 0. \label{eq:local}
\end{equation}
When this is satisfied, there is no dynamics generated locally by oscillations. Neutrino--neutrino forward scattering contributes a term to $H_{\boldsymbol{p}} (t, \boldsymbol{x})$ that couples $\rho_{\boldsymbol{p}} (t, \boldsymbol{x})$ to $\rho_{\boldsymbol{q}} (t, \boldsymbol{x})$ for all momenta $\boldsymbol{q}$. We let $H^\textrm{eq}_{\boldsymbol{p}} (t, \boldsymbol{x})$ signify the Hamiltonian evaluated at $\rho_{\boldsymbol{q}} (t, \boldsymbol{x})= \rho^\textrm{eq}_{\boldsymbol{q}} (t, \boldsymbol{x})$, again for all momenta. For any given $(t, \boldsymbol{x})$, Eq.~\eqref{eq:local} represents a system of coupled equations, one for each momentum $\boldsymbol{p}$.

Determining $\rho^\textrm{eq}_{\boldsymbol{p}}$ and $H^\textrm{eq}_{\boldsymbol{p}}$ from Eq.~\eqref{eq:local} is equivalent to solving the self-consistency conditions of Ref.~\cite{johns2023c}. The instantaneous-relaxation treatment arrives at the same procedure and therefore the same $\rho_{\boldsymbol{p}}^\textrm{eq}$.

Let us assume for simplicity that there are two neutrino flavors. Then we can rewrite Eq.~\eqref{eq:local} by expanding $\rho^\textrm{eq}_{\boldsymbol{p}}$ and $H^\textrm{eq}_{\boldsymbol{p}}$ in the basis of Pauli matrices $\sigma_i$:
\begin{align}
&\rho^\textrm{eq}_{\boldsymbol{p}} (t, \boldsymbol{x}) = \frac{1}{2} \left( P^\textrm{eq}_{\boldsymbol{p},0} (t, \boldsymbol{x}) \mathbbm{1} + \boldsymbol{P}^\textrm{eq}_{\boldsymbol{p}} (t, \boldsymbol{x}) \cdot \boldsymbol{\sigma} \right), \notag \\
&H^\textrm{eq}_{\boldsymbol{p}} (t, \boldsymbol{x}) = \frac{1}{2} \left( H^\textrm{eq}_{\boldsymbol{p},0} (t, \boldsymbol{x}) \mathbbm{1} + \boldsymbol{H}^\textrm{eq}_{\boldsymbol{p}} (t, \boldsymbol{x}) \cdot \boldsymbol{\sigma} \right).
\end{align}
The commutator of the parts proportional to the identity matrix vanishes trivially. The nontrivial part is
\begin{equation}
\boldsymbol{H}^\textrm{eq}_{\boldsymbol{p}} (t, \boldsymbol{x}) \times \boldsymbol{P}^\textrm{eq}_{\boldsymbol{p}} (t, \boldsymbol{x}) = 0. \label{eq:cross}
\end{equation}
We can therefore write the formal solution
\begin{equation}
\boldsymbol{P}^\textrm{eq}_{\boldsymbol{p}} (t, \boldsymbol{x}) = s_{\boldsymbol{p}} (t, \boldsymbol{x}) P^\textrm{eq}_{\boldsymbol{p}} (t, \boldsymbol{x}) \boldsymbol{\hat{H}}^\textrm{eq}_{\boldsymbol{p}} (t, \boldsymbol{x}) \label{eq:adiabtx}
\end{equation}
with alignment factor
\begin{equation}
s_{\boldsymbol{p}} (t, \boldsymbol{x}) = \boldsymbol{\hat{H}}^\textrm{eq}_{\boldsymbol{p}} (t, \boldsymbol{x}) \cdot \boldsymbol{\hat{P}}^\textrm{eq}_{\boldsymbol{p}} (t, \boldsymbol{x}).
\end{equation}
Note that the slow (astrophysical-scale) variation of $\rho^\textrm{eq}_{\boldsymbol{p}} (t, \boldsymbol{x})$ implies that $\boldsymbol{\hat{H}}^\textrm{eq}_{\boldsymbol{p}} (t, \boldsymbol{x})$ is constant on oscillation scales. To obtain an explicit solution, the values of $P^\textrm{eq}_{\boldsymbol{p}} (t, \boldsymbol{x})$ and $s_{\boldsymbol{p}} (t, \boldsymbol{x})$ must be known. To completely specify the neutrino distributions, $P^\textrm{eq}_{\boldsymbol{p},0} (t, \boldsymbol{x})$ must be known as well. These are all set globally, \textit{i.e.}, by solving Eq.~\eqref{eq:notmisc} throughout the history and extent of the astrophysical system.

For the purposes of simulating a supernova or merger, we want a coarse-grained formulation of neutrino transport. That is, we want to incorporate flavor mixing into transport without having to resolve scales much smaller than $l_\textrm{astro}$. Assuming a resolution on the order of $l_\textrm{astro}$ and using Eq.~\eqref{eq:lastro}, we see that $\rho_{\boldsymbol{p}} (t, \boldsymbol{x})$ is approximately homogeneous on subgrid scales. Following the Appendix of Ref.~\cite{johns2023c}, we define the coarse-graining operator
\begin{equation}
\langle \cdot \rangle (t, \boldsymbol{x}) = \frac{1}{V \Delta t} \int_{\mathcal{R}_{\boldsymbol{x}}} d^3 \boldsymbol{x}' \int_t^{t+\Delta t} dt' ~ (\cdot) (t', \boldsymbol{x}') \label{eq:cgop}
\end{equation}
over time interval $\Delta t$ and region $\mathcal{R}_{\boldsymbol{x}}$ centered at $\boldsymbol{x}$. Let $\Delta t$ and the volume $V$ of region $\mathcal{R}_{\boldsymbol{x}}$ be set by the simulation step size and grid resolution. Then
\begin{equation}
\langle \rho^\textrm{eq}_{\boldsymbol{p}} \rangle (t, \boldsymbol{x}) \cong \rho^\textrm{eq}_{\boldsymbol{p}} (t, \boldsymbol{x}), \label{eq:cghomog}
\end{equation}
ignoring the small deviation from equality that comes from the use of $\Delta t$ and $V$ at the threshold scale where inhomogeneity begins to emerge. Using a coarse-graining relevant to simulations therefore leaves Eq.~\eqref{eq:notmisc} approximately unchanged.

The miscidynamic equation is derived in Ref.~\cite{johns2023c} by coarse-graining neutrino transport and using thermodynamic principles. Here we have reproduced this result by self-consistently applying the notion of instantaneous relaxation to a local steady state [Eq.~\eqref{eq:tau}]. As described in Sec.~\ref{sec:issue}, this is the ansatz that asymptotic-state subgrid models implement inconsistently.

\section{Adiabaticity\label{sec:adiab}}

The instantaneous-relaxation ansatz of the previous sections is tantamount to the adiabatic approximation. We explain this point below.

\subsection{Quantum-mechanical adiabaticity}

Adiabaticity has been a central concept in neutrino physics since the discovery of the MSW effect \cite{mikheyev1985, bethe1986, messiah1986b}. In that particular application, adiabaticity entails no more than the standard quantum-mechanical concept referenced in the adiabatic theorem. Consider a neutrino produced in the center of the Sun whose initial state $| \psi (0) \rangle$ in flavor Hilbert space has overlap $| \langle \nu (0) | \psi (0) \rangle |$ with a certain instantaneous energy eigenstate $| \nu (0) \rangle$. If the flavor evolution of the neutrino is adiabatic, then the overlap remains the same,
\begin{equation}
| \langle \nu (t) | \psi (t) \rangle | = \textrm{constant}, \label{eq:msw}
\end{equation}
as the neutrino travels through the solar medium.

The scope of the adiabaticity concept needs to be broadened to make it suitable to flavor evolution in compact objects. Neutrino--neutrino forward scattering gives rise to terms in $H_{\boldsymbol{p}} (t, \boldsymbol{x})$ that depend on $\rho_{\boldsymbol{q}} (t, \boldsymbol{x})$. For the moment, assume homogeneity (no $\boldsymbol{x}$-dependence) and collisionless evolution ($iC_{\boldsymbol{p}} = 0$). Then the quantum kinetic equation reduces to
\begin{equation}
i \frac{d}{dt} \rho_{\boldsymbol{p}} (t) = \left[ H_{\boldsymbol{p}} (t), \rho_{\boldsymbol{p}} (t) \right]. \label{eq:rsqke}
\end{equation}
The setting is nonlinear mean-field quantum mechanics. Accordingly, we move from the Hilbert space in which state $| \psi (t) \rangle$ resides to the flavor space containing polarization vector $\boldsymbol{P}_{\boldsymbol{p}}(t)$. Adiabaticity can then be defined by the condition
\begin{equation}
\boldsymbol{\hat{H}}_{\boldsymbol{p}} (t) \cdot \boldsymbol{\hat{P}}_{\boldsymbol{p}} (t) = \textrm{constant} \label{eq:adiabt}
\end{equation}
or, if the initial ensemble consists of energy eigenstates,
\begin{equation}
\boldsymbol{P}_{\boldsymbol{p}} (t) = s P_{\boldsymbol{p}} (t) \boldsymbol{\hat{H}}_{\boldsymbol{p}} (t) \label{eq:adiabt2}
\end{equation}
with constant $s = \pm 1$. This encompasses externally prescribed time-dependent Hamiltonians as well as $\rho$-dependent ones, for which the energy eigenstates depend on the state of the system. Note the similarity of Eq.~\eqref{eq:adiabt2} to Eq.~\eqref{eq:adiabtx}. The former is a statement of adiabaticity. The latter expresses local self-consistency of the relaxation approximation of neutrino quantum kinetics.

There is a precedent for self-consistency conditions in simple models of collective oscillations. Duan, Fuller, Carlson, and Qian highlighted adiabaticity as an important analytic concept underpinning their influential calculations of flavor transformation in supernovae \cite{duan2006, duan2006b, duan2006c, duan2007, duan2007b, duan2007c}. A thorough analysis of the phenomenon of spectral swaps/splits was then presented by Raffelt and Smirnov \cite{raffelt2007, raffelt2007c}, who obtained explicit adiabatic solutions $\boldsymbol{P}_{\boldsymbol{p}} (r)$ by solving Eq.~\eqref{eq:adiabt} self-consistently. (Stationarity of the flavor field is assumed and $t$ is replaced by radius $r$.) The self-consistency conditions derived in Refs.~\cite{raffelt2007, raffelt2007c} ensure that Eq.~\eqref{eq:adiabt} is satisfied for all momenta simultaneously.

Raffelt--Smirnov adiabaticity addresses local self-consistency but sidesteps the issue of global self-consistency. Imagine that Eq.~\eqref{eq:rsqke} (with $t \rightarrow r$) describes the flavor evolution outward in radius. The development of $\boldsymbol{P}_{\boldsymbol{p}} (r)$ is fully prescribed by boundary conditions at the innermost radius. Advection and collisions, which relate adjacent regions to each other nontrivially, are effectively absent.

\subsection{Thermodynamic adiabaticity}

Miscidynamics generalizes adiabaticity further. It makes the concept applicable to quantum kinetics, including advection ($\partial_t + \boldsymbol{\hat{p}} \cdot \partial_{\boldsymbol{x}}$) and collisions ($iC_{\boldsymbol{p}}$), and imposes both local and global self-consistency.

In the quantum-kinetic setting, $\rho^\textrm{eq}_{\boldsymbol{p}} (t, \boldsymbol{x})$ at some particular $(t, \boldsymbol{x})$ depends on the whole history of the supernova or merger. This is exceptional in the context of quantum mechanics, where adiabatic evolution is typically a function of an external control parameter. For example, if a spin adiabatically follows an applied magnetic field, it is only necessary to know the instantaneous field direction to know the spin direction at the same moment. Similarly, a Raffelt--Smirnov adiabatic solution is essentially determined at time $t$ by the neutrino self-interaction parameter $\mu (t)$ \cite{raffelt2007, raffelt2007c}. Global dependence is less exceptional in the context of classical statistical physics. Notably, the same property is true of fluid flows. In Sec.~\ref{sec:non} we sketch some of the parallels between miscidynamics and hydrodynamics.

The generalization of adiabaticity uses neutrino quantum thermodynamics, wherein each coarse-grained $\langle \rho_{\boldsymbol{p}} \rangle (t, \boldsymbol{x})$ is associated with an ensemble of microstates $\rho_{\boldsymbol{p}} (t, \boldsymbol{x})$ defined over the same coarse-graining region $\mathcal{R}_{\boldsymbol{x}}$. The ensemble allows for the definition of an equilibrium state via
\begin{equation}
\rho^\textrm{eq}_{\boldsymbol{p}} (t, \boldsymbol{x}) \equiv \overline{\rho_{\boldsymbol{p}}} (t, \boldsymbol{x}), \label{eq:ergeq}
\end{equation}
where the quantity on the right-hand side is the ensemble expectation value, which can be calculated by maximizing a coarse-grained entropy. The approximation
\begin{equation}
\langle \rho_{\boldsymbol{p}} \rangle (t, \boldsymbol{x}) \cong \overline{\rho_{\boldsymbol{p}}} (t, \boldsymbol{x}) \label{eq:ergline}
\end{equation}
is the ergodic hypothesis applied to neutrino oscillations. The miscidynamic equation in the adiabatic limit [Eq.~\eqref{eq:misc}] can be derived by assuming that ergodicity is a good approximation even on infinitesimally small scales. With neutrinos always in local equilibrium, oscillations lead to no heat production (\textit{i.e.}, flavor mixing is adiabatic in the thermodynamic sense), as can be confirmed using the definitions in Ref.~\cite{johns2023c}. The pathway through thermodynamic adiabaticity is an alternative to the one in Sec.~\ref{sec:misc}, where miscidynamics was arrived at by assuming instantaneous relaxation to a homogeneous asymptotic state.

When adiabatic miscidynamics is applied to the simple model of Refs.~\cite{raffelt2007, raffelt2007c}, it reduces to Raffelt--Smirnov adiabaticity. Computationally, it is encouraging that these studies have already demonstrated the feasibility of solving self-consistency conditions of this type.

\subsection{Classical-mechanical adiabaticity}

The expanded adiabaticity concept unifies quantum-mechanical, thermodynamic, and classical-mechanical notions of adiabaticity. The connection to classical mechanics is made through the slow \cite{hannestad2006, duan2007b, johns2018}, fast \cite{johns2020, padillagay2022, fiorillo2023}, and collisional \cite{johns2023d} neutrino flavor pendula. The last of these works addresses neutrino quantum thermodynamics explicitly.

\subsection{Microscopic and macroscopic}

Paradigms \textbf{(2)} and \textbf{(3)} present different visions of what transpires in a supernova or merger. In short, \textbf{(2)} imagines that flavor instabilities occur abundantly in these sites, whereas \textbf{(3)} imagines that neutrino flavor distributions adiabatically adjust to the spatial and temporal variation of local mixing equilibrium. Adiabatic adjustment can be understood from a microscopic and a macroscopic viewpoint, as explained below.

The project initiated in Ref.~\cite{johns2023c} is the development of a statistical theory of neutrino oscillations. That study principally concentrated on the equilibrium part of the theory. By formulating miscidynamics in the adiabatic limit, it did not need to quantitatively confront the issues of nonequilibrium dynamics and unstable equilibria. These parts of the theory are currently less developed.

The relaxation-time approximation of Ref.~\cite{nagakura2023d} proposes that the nonequilibrium dynamics can be modeled using
\begin{equation}
\left[ H_{\boldsymbol{p}} (t, \boldsymbol{x}), \rho_{\boldsymbol{p}} (t, \boldsymbol{x}) \right] \cong \gamma (t, \boldsymbol{x}) \left( \rho^\textrm{eq}_{\boldsymbol{p}} (t, \boldsymbol{x}) - \rho_{\boldsymbol{p}} (t, \boldsymbol{x}) \right). \label{eq:rta1}
\end{equation}
This has a precursor in the proposal of Ref.~\cite{johns2019b}, which applies the relaxation-time approximation to oscillations and collisions simultaneously for the purposes of describing the production of sterile neutrino dark matter in the early universe. Coarse-graining the collisional dynamics is useful for that application, unlike for the oscillation problem in compact objects. Note that Eq.~\eqref{eq:rta1} is fundamentally distinct from the application of the relaxation-time approximation to collisional processes alone. For discussion of the latter in the context of neutrino mixing, see Ref.~\cite{hannestad2015}.

Equation~\eqref{eq:rta1} illustrates the microscopic view of adiabatic adjustment. At each moment and location, collisions, advection, and the evolution of the astrophysical fluid all drive neutrinos away from local mixing equilibrium. In response to these disturbances, subgrid neutrino oscillations bring the system back toward equilibrium. The numerical value of $\gamma$ is thus tied to the linear response near local mixing equilibrium. At present, this picture is mostly conceptual. Further development of the statistical theory of neutrino oscillations could potentially establish a microphysical basis for Eq.~\eqref{eq:rta1} by deriving it from neutrino quantum kinetics. This would make it possible to calculate $\gamma$ rather than merely estimate it, and would provide a more complete microscopic picture of how neutrinos attempt to maintain local mixing equilibrium.

In the adiabatic limit, subgrid oscillations are flawlessly effective at maintaining equilibrium. Microscopically, nonadiabaticity arises from the finite time scale of the flavor-mixing response. We elaborate on the issue of nonadiabaticity in Sec.~\ref{sec:non}.

The macroscopic view of adiabatic adjustment is not based on oscillations per se---those are microscopic processes according to neutrino quantum thermodynamics---but rather on ergodicity. In a thermodynamic ensemble, the most probable state is $\rho^\textrm{eq}_{\boldsymbol{p}}$. If ergodicity holds over arbitrarily small scales, then the coarse-grained flavor evolution is equal to the most probable flavor evolution: adiabatic tracking of local mixing equilibrium [Eqs.~\eqref{eq:ergeq} and \eqref{eq:ergline}]. Nonadiabaticity is due to the fact that in reality ergodicity does not apply over arbitrarily short distances and time spans.

The microscopic and macroscopic viewpoints are of course compatible with one another. Oscillations are responsible for flavor-space ergodicity.

It is important to stress that ergodicity is only a hypothesis. Quantitative analyses asking whether collective oscillations are chaotic have come to positive conclusions \cite{hansen2014, urquilla2024}. Qualitatively, the existence of fluctuating quasi-steady states is consistent with ergodic evolution. Moreover, in the systems that appear to be obviously nonergodic---the slow and fast flavor pendula---it is understood why: these systems are rendered integrable by symmetries and conservation laws. Further investigation of the neutrino ergodic hypothesis is warranted. If ergodicity proves to be a useful approximation, the essential question will be to determine the manifold of flavor states on which it is applicable.

\section{Nonadiabaticity\label{sec:non}}

Miscidynamics is the local-equilibrium theory of \textit{oscillating} particles. Hydrodynamics is the local-equilibrium theory of \textit{colliding} particles. Though paradigm \textbf{(3)} is a significant departure from other approaches to the oscillation problem, it closely parallels the hydrodynamic paradigm. The comparison of the two illuminates a possible path leading to nonadiabatic miscidynamics.

\subsection{Characteristic numbers}

Up to this point we have assumed that flavor relaxation is instantaneous: $\tau \cong 0$. To allow for nonadiabaticity (\textit{i.e.}, some degree of deviation from local mixing equilibrium), we must consider finite $\tau$.

Roughly speaking, relaxation occurs on the time scale of oscillations because it is driven by the flavor-mixing Hamiltonian $H_{\boldsymbol{p}}$. Let us define a dimensionless number
\begin{equation}
\varepsilon \equiv \frac{l_\textrm{osc}}{l_\textrm{astro}},
\end{equation}
the ratio of typical oscillation and astrophysical length scales. We expect this characteristic parameter to indicate the importance of nonadiabaticity in miscidynamic evolution. Adiabatic miscidynamics [Eq.~\eqref{eq:misc}] assumes $\varepsilon \cong 0$. Nonadiabatic terms must be added to Eq.~\eqref{eq:misc} if finite-$\varepsilon$ effects are thought to be significant. For $\varepsilon \gtrsim 1$, miscidynamics is not expected to be a good approximation of quantum kinetics.

The analogous parameter in hydrodynamics is the Knudsen number
\begin{equation}
\textrm{Kn} \equiv \frac{l_\textrm{mfp}}{l_\textrm{scale}},
\end{equation}
the ratio of collisional mean free path to the relevant length scale of the problem. Roughly speaking, $\textrm{Kn}$ quantifies the importance of viscous effects. Fluid descriptions are questionable for $\textrm{Kn} \gtrsim 1$. This is in fact the regime that applies to oscillating astrophysical neutrinos, which is why we propose using a near-equilibrium treatment of flavor but not momentum transport.

These dimensionless numbers, $\varepsilon$ and $\textrm{Kn}$, are merely suggestive of the qualitative behavior. Their precise values are somewhat ambiguous because a system is not uniquely specified by, for example, a single mean free path and a single macroscopic scale.

We emphasize that $\varepsilon$ is distinct from the expansion parameters ($p \equiv | \boldsymbol{p} |$)
\begin{equation}
\frac{\partial_{t,\boldsymbol{x}}}{p}, ~ \frac{m_\nu}{p}, ~ \frac{l_\textrm{osc}^{-1}}{p} \label{eq:qkeparams}
\end{equation}
that are used to derive quantum kinetics from quantum field theory \cite{sigl1993, vlasenko2014}. Expanding in these small quantities effectively coarse-grains over the quantum (de Broglie) microscale. Miscidynamics is based on a further coarse-graining over the oscillation mesoscale \cite{johns2023c}, hence the introduction of the new parameter $\varepsilon$.

\subsection{The miscidynamics/hydrodynamics analogy}

Miscidynamics describes kinetic systems that are near local mixing equilibrium. Hydrodynamics describes kinetic systems that are near local collisional equilibrium.

Fluid-like behavior arises at small $\textrm{Kn}$. In this regime, particle collisions are frequent enough that they can be treated as unresolved, near-equilibrium microphysics. Discreteness---the individuality of particles and scattering events---is only evident in the continuum flow through viscosity.

For an incompressible fluid in the absence of external forces, the velocity field $\boldsymbol{v} (t, \boldsymbol{x})$ obeys the Navier--Stokes equation
\begin{equation}
\left( \partial_t + \boldsymbol{v} \cdot \partial_{\boldsymbol{x}} \right) \boldsymbol{v} (t, \boldsymbol{x}) = - \frac{\partial_{\boldsymbol{x}} \pi (t, \boldsymbol{x})}{\varrho (t, \boldsymbol{x})} + \nu \partial^2_{\boldsymbol{x}} \boldsymbol{v} (t, \boldsymbol{x}), \label{eq:ns}
\end{equation}
where $\varrho$ is the mass density, $\pi$ is the pressure, and $\nu$ is the kinematic viscosity. As $\lambda_\textrm{mfp} \rightarrow 0$ ($\textrm{Kn} \rightarrow 0$), local collisional relaxation becomes instantaneous and viscous effects become negligible.

The analogue of adiabatic miscidynamics is the inviscid ($\nu \cong 0$) Navier--Stokes equation, \textit{i.e.}, the Euler equation. Transport under the Euler equation is adiabatic with respect to collisions. Particles have infinitesimal mean free paths and are unable to separate from each other for long enough to reach significantly different regions of the fluid. Although $\boldsymbol{v} (t, \boldsymbol{x})$ has nonzero spatial gradients, diffusion is negligible because particles are constantly exchanging momentum with the macroscopic flow.

For $\nu \neq 0$, there is a finite scale below which fluid kinetic energy is dissipated via diffusion. On small enough scales, particles travel freely from the flow and exhibit deviations from the local equilibrium momentum distribution. The competition between advection (which transports particles between regions of different macroscopic properties) and collisions (which bring particles into local equilibrium) generates heat.

Viscosity approximates relaxation due to unresolved microscopic interactions. Fundamentally, irreversibility comes from molecular chaos: particles effectively decorrelate in between interactions. 

The analogy with hydrodynamics points the way toward further development of miscidynamics. The extension of the $\mathcal{O} (\textrm{Kn}^0)$ Euler equation to the $\mathcal{O} (\textrm{Kn}^1)$ Navier--Stokes equation can be paralleled by the formulation of $\mathcal{O} (\varepsilon^1)$ miscidynamics.

\subsection{Nonadiabatic miscidynamics}

Irreversibility arises from decorrelation in neutrino oscillations as well. Nonadiabatic corrections to miscidynamics approximate the effects on the coarse-grained evolution of this information loss.

Consider the application of the coarse-graining operator [Eq.~\eqref{eq:cgop}] to the quantum kinetic equation [Eq.~\eqref{eq:qke}] in the local equilibrium frame of Sec.~\ref{sec:misc}:
\begin{align}
i \left( \partial_t + \boldsymbol{\hat{p}} \cdot \partial_{\boldsymbol{x}} \right) \left\langle \rho_{\boldsymbol{p}} \right\rangle (t, \boldsymbol{x}) = &\left\langle \left[ H_{\boldsymbol{p}}, \rho_{\boldsymbol{p}} \right] \right\rangle (t, \boldsymbol{x}) \notag \\
&~~~~~ + \left\langle i C_{\boldsymbol{p}} \right\rangle (t, \boldsymbol{x}). \label{eq:cgqke}
\end{align}
The derivatives are moved outside of the coarse-graining operator following the derivation in the Appendix of Ref.~\cite{johns2023c}. Next we introduce $\delta \rho_{\boldsymbol{p}} (t, \boldsymbol{x})$, the fine-grained deviation from the average, by writing
\begin{equation}
\rho_{\boldsymbol{p}} (t, \boldsymbol{x}) = \left\langle \rho_{\boldsymbol{p}} \right\rangle (t, \boldsymbol{x}) + \delta \rho_{\boldsymbol{p}} (t, \boldsymbol{x}). \label{eq:deltarho}
\end{equation}
The right-hand side of Eq.~\eqref{eq:cgqke} contains correlations with structures like
\begin{equation}
\langle \delta \rho_{\boldsymbol{q}} \delta \rho_{\boldsymbol{p}} \rangle (t, \boldsymbol{x}).
\end{equation}
Correlations with more factors of $\delta \rho$ appear due to Pauli blocking in the collision integrals. We assume that all of these terms are small in magnitude and drop the ones that appear in $\langle i C_{\boldsymbol{p}} \rangle$, since the collision term is already at order $G_F^2$.

The coarse-grained commutator in Eq.~\eqref{eq:cgqke} can be split into uncorrelated and correlated parts:
\begin{equation}
\left\langle \left[ H_{\boldsymbol{p}}, \rho_{\boldsymbol{p}} \right] \right\rangle (t, \boldsymbol{x}) = \left[  \left\langle H_{\boldsymbol{p}} \right\rangle (t, \boldsymbol{x}), \left\langle \rho_{\boldsymbol{p}} \right\rangle (t, \boldsymbol{x}) \right] + h_{\boldsymbol{p}} (t, \boldsymbol{x}),
\end{equation}
where $\left\langle H_{\boldsymbol{p}} \right\rangle (t, \boldsymbol{x})$ is defined to be $H_{\boldsymbol{p}} (t, \boldsymbol{x})$ evaluated using $\rho_{\boldsymbol{q}}  (t, \boldsymbol{x}) \rightarrow \langle \rho_{\boldsymbol{q}} \rangle  (t, \boldsymbol{x})$. The correlated part $h_{\boldsymbol{p}} (t, \boldsymbol{x})$ is a functional of $\langle [ \rho_{\boldsymbol{q}}, \rho_{\boldsymbol{p}}] \rangle  (t, \boldsymbol{x})$.

Assume that $\rho_{\boldsymbol{p}}$ fluctuates around local mixing equilibrium: $\langle \rho_{\boldsymbol{p}} \rangle (t, \boldsymbol{x}) = \rho^\textrm{eq}_{\boldsymbol{p}} (t, \boldsymbol{x})$. We then have
\begin{equation}
\left[ \left\langle H_{\boldsymbol{p}} \right\rangle (t, \boldsymbol{x}), \left\langle \rho_{\boldsymbol{p}} \right\rangle (t, \boldsymbol{x}) \right] = \left[ H^\textrm{eq}_{\boldsymbol{p}} (t, \boldsymbol{x}), \rho^\textrm{eq}_{\boldsymbol{p}} (t, \boldsymbol{x}) \right] = 0. 
\end{equation}
Compared to the local self-consistency condition in Eq.~\eqref{eq:local}, the difference is that here we impose local equilibrium only at the coarse-grained level. Taking local equilibrium to hold down to arbitrarily small scales leads to adiabatic miscidynamics. By allowing for a scale below which nonequilibrium fluctuations are significant, we permit nonadiabaticity in the coarse-grained dynamics.

Applying the logic of Sec.~\ref{sec:misc} and the steps outlined in this subsection to Eq.~\eqref{eq:cgqke}, we obtain the transport equation
\begin{equation}
i \left( \partial_t + \boldsymbol{\hat{p}} \cdot \partial_{\boldsymbol{x}} \right) \rho^\textrm{eq}_{\boldsymbol{p}} (t, \boldsymbol{x}) = h_{\boldsymbol{p}} (t, \boldsymbol{x}) + i C^\textrm{eq}_{\boldsymbol{p},\textrm{non}} (t, \boldsymbol{x}), \label{eq:nonmisc}
\end{equation}
which describes nonadiabatic miscidynamics. To be useful it must be supplemented with a specific model for $h_{\boldsymbol{p}} (t, \boldsymbol{x})$. Ultimately the choice should be motivated by the microphysics. As suggested in Ref.~\cite{johns2023c}, one could attempt to calculate the miscidynamic analogue of viscosity using the analogue of molecular chaos.

It is unclear whether nonadiabaticity is practically important in supernovae and mergers. In general we expect $\varepsilon \ll 1$ because oscillation lengths are typically much smaller than the collisional mean free path and the scale over which the astrophysical medium varies. Shocks are obvious exceptions, but a sharp change in the ambient matter properties may not be sufficient to drive $\rho_{\boldsymbol{p}}$ far from equilibrium.

\subsection{Subgrid fluctuations}

From Eq.~\eqref{eq:deltarho} we have
\begin{equation}
\left\langle \rho_{\boldsymbol{q}} \rho_{\boldsymbol{p}} \right\rangle = \left\langle \rho_{\boldsymbol{q}} \right\rangle \left\langle \rho_{\boldsymbol{p}} \right\rangle + \left\langle \delta\rho_{\boldsymbol{q}} \delta\rho_{\boldsymbol{p}} \right\rangle, \label{eq:cgprod}
\end{equation}
with all coarse-grained averages being functions of the same $(t, \boldsymbol{x})$. Ergodicity relates the quantities in Eq.~\eqref{eq:cgprod} to ensemble expectation values [Eq.~\eqref{eq:ergline}]. At equilibrium,
\begin{equation}
\left\langle \rho_{\boldsymbol{q}} \rho_{\boldsymbol{p}} \right\rangle = \left( \rho_{\boldsymbol{q}} \rho_{\boldsymbol{p}} \right)^\textrm{eq} = \rho_{\boldsymbol{q}}^\textrm{eq} \rho_{\boldsymbol{p}}^\textrm{eq} = \left\langle \rho_{\boldsymbol{q}} \right\rangle \left\langle \rho_{\boldsymbol{p}} \right\rangle. \label{eq:factor}
\end{equation}
The factorization in Eq.~\eqref{eq:factor} can be understood as the statement that thermal fluctuations around equilibrium become vanishingly small relative to mean values in the thermodynamic limit.

We can also consider what Eq.~\eqref{eq:factor} says about the fine-grained kinetics. Factorization occurs when subgrid fluctuations are uncorrelated or simply vanish. If the flavor evolution in an astrophysical event is adiabatic, then fluctuations are not generated at all by oscillations. Under Eq.~\eqref{eq:misc}, neutrino distributions are constant and homogeneous on subgrid scales, always being in local mixing equilibrium.

Fast flavor instabilities are highly nonadiabatic. They generate small-scale inhomogeneity and temporal fluctuations. Late-time asymptotic states may nonetheless be instances of mixing equilibrium \cite{johns2023c}. In instability calculations, correlations $h_{\boldsymbol{p}}$ are anticipated to be small because the subgrid fluctuations, though not small themselves, are uncorrelated.

Recently it was noticed that kinetic energy (associated with neutrino momentum $\boldsymbol{p}$) and internal energy (associated with flavor polarization $\boldsymbol{P}$) are not independently conserved under quantum-kinetic evolution \cite{fiorillo2024}. Inhomogeneity in $\boldsymbol{P}$ mediates an exchange of the two types of energy. Ref.~\cite{fiorillo2024} numerically demonstrated how inhomogeneity generated by fast flavor instability causes significant nonconservation of internal energy.

The coupling of the internal energy to the much larger kinetic energy raises a potential concern for miscidynamics: Is adiabatic flavor evolution even possible given that neutrino advection can act as an energetic source or sink for the flavor-mixing sector? 

Upon inspection it becomes clear that the kinetic--internal coupling is not a problem for adiabatic miscidynamics after all. The exchange of energy is mediated by spatial gradients of $\rho_{\boldsymbol{p}} (t, \boldsymbol{x})$ \cite{fiorillo2024}. But recall from Eq.~\eqref{eq:lastro} that gradients are on the order of $l_\textrm{astro}^{-1}$. The $\varepsilon \rightarrow 0$ limit has $l_\textrm{astro} \rightarrow \infty$, so that
\begin{equation}
\partial_{\boldsymbol{x}} \rho_{\boldsymbol{p}} (t,\boldsymbol{x}) \longrightarrow 0
\end{equation}
in the adiabatic limit. This is yet another form of self-consistency satisfied by miscidynamics. In the adiabatic limit, flavor is homogeneous. In the homogeneous limit, nonconservative terms vanish.

It is also possible for conservative coarse-grained motion to emerge from the averaging-out of nonconservative fine-grained fluctuations \cite{johns2023d}. Indeed, neutrino quantum thermodynamics does not require the parameters characterizing an ensemble to be strictly constant. The requirement is only that they have fixed expectation values. Fluctuations are permissible and expected.

\section{Summary \label{sec:summary}}

This work has made three main points. First, asymptotic-state subgrid models inconsistently impose local flavor relaxation (Sec.~\ref{sec:issue}). Second, instantaneous relaxation to homogeneous asymptotic states implies adiabatic miscidynamics (Secs.~\ref{sec:misc} and \ref{sec:adiab}). Third, the extension of the Euler equation to the Navier--Stokes equation is a conceptual road map for the development of nonadiabatic miscidynamics (Sec.~\ref{sec:non}).

The implementation of adiabatic miscidynamics into astrophysical simulations is briefly outlined in Ref.~\cite{johns2023c}. It has not yet been carried out in practice. Further work is also needed to formulate nonadiabatic miscidynamics in detail.

\begin{acknowledgments}
The author gratefully acknowledges conversations with Damiano Fiorillo, Chris Fryer, Jonah Miller, Hiroki Nagakura, and Georg Raffelt. This work was supported by a Feynman Fellowship through LANL LDRD project number 20230788PRD1.
\end{acknowledgments}

\bibliography{all_papers}

\end{document}